\documentclass[twocolumn,showpacs,preprintnumbers,amsmath,amssymb]{revtex4}

\usepackage{graphicx}
\usepackage{dcolumn}
\usepackage{bm}

\begin{document}

\preprint{}

\title{On energy and momentum of an ultrarelativistic unstable system}

\author{Gianluca Geloni}
\email{g.a.geloni@tue.nl}

\affiliation{%
Department of Applied Physics, Technische Universiteit Eindhoven,
\\
 P.O. Box 513, 5600MB Eindhoven, The Netherlands
}%

\author{Evgeni Saldin}
\affiliation{Deutsches Elektronen-Synchrotron DESY, \\
Notkestrasse 85, 22607 Hamburg, Germany
}%

\begin{abstract}
We study an electron bunch together with its self-fields from the
viewpoint of basic dynamical quantities. This leads to a
methodological discussion about the definition of energy and
momentum for fully electromagnetic systems and about the relation
between covariance of the energy-momentum pair and stability. We
show here that, in the case of unstable systems, there is no mean
to define, in a physically meaningful way, a total energy-momentum
four-vector: covariance of the energy-momentum pair follows from
the stability of the system and viceversa, as originally pointed
out by Henry Poincar\'e.
\end{abstract}


\begin{widetext}
\thispagestyle{empty}
\begin{large}
\textbf{DEUTSCHES ELEKTRONEN-SYNCHROTRON}\\
\end{large}

DESY 02-201  

November 2002

\begin{eqnarray}
\nonumber &&\cr \nonumber && \cr \nonumber &&\cr
\end{eqnarray}

\begin{center}
\begin{Large}
\textbf{On energy and momentum of an ultrarelativistic unstable
system}
\end{Large}
\begin{eqnarray}
\nonumber &&\cr \nonumber && \cr
\end{eqnarray}

\begin{large}
Gianluca Geloni
\end{large}

\textsl{\\Department of Applied Physics, Technische Universiteit
Eindhoven, \\P.O. Box 513, 5600MB Eindhoven, The Netherlands}
\begin{eqnarray}
\nonumber
\end{eqnarray}
\begin{large}
Evgeni Saldin
\end{large}

\textsl{\\Deutsches Elektronen-Synchrotron DESY, \\Notkestrasse
85, 22607 Hamburg, Germany}
\begin{eqnarray}
\nonumber
\end{eqnarray}

\newpage

\end{center}
\end{widetext}

\maketitle

\maketitle

\section{\label{INTRO}INTRODUCTION}

Nearly one hundred years have passed since Abraham and Lorentz
calculated their famous expressions for the energy and momentum of
a purely electromagnetic, spherically symmetrical distribution of
charges \cite{ABR1}, \cite{LORE}. This distribution constitutes an
attempt to build a classical model of the electron: according to
Lorentz's initial idea, mass, energy and momentum of the electron
could, indeed, be of completely elecromagnetic nature.

Nevertheless, the energy (divided by the speed of light in vacuum,
as we will understand through this paper) and momentum of such an
electromagnetic electron do not constitute a four-vector. In fact
(as Abraham \cite{ABR2} pointed out already in 1904 probably, at
that time, without a clear understanding of what a four-vector
is), in a frame moving with velocity ${\bm v}$ with respect to the
system rest frame, we have

\begin{equation}
E_{e}  = \gamma U'(1 + 1/3 \beta^2) \label{AL2}
\end{equation}
and

\begin{equation}
{\bm P_{e}} = 4/3 \gamma {\bm v} U'/c^2, \label{AL1}
\end{equation}
where the index $e$ indicates the electromagnetic nature of the
energy $E_{e}$ and momentum ${\bm P_{e}}$, $\gamma$ is the usual
Lorentz factor, $\beta$ is the velocity $v/c$ (normalized to the
speed of light in vacuum $c$), and $U'$ indicates the
electromagnetic energy in the electron rest frame \cite{JACK},

\begin{equation}
U' = \epsilon_{0}/2 \int {\bm E'}^2 dV', \label{uprimu}
\end{equation}
where $\epsilon_{0}$ is the free space permittivity. $U'$ is
purely an electrostatic quantity (in this paper the prime will
always indicate quantities calculated in the rest frame; therefore
${\bm E'}$ and $dV'$ are, respectively, the electric field and the
volume element in the rest frame of the system).

It is worth to mention here that the factor $4/3$ in Eq.
(\ref{AL1}) and the term proportional to $1/3 \beta^2$ in Eq.
(\ref{AL2}) depend on the choice of spherical symmetry made on the
charge distribution: had we chosen, for instance, an infinitely
long line distribution in the direction of ${\bm v}$, we would
have found

\begin{equation}
E_{e2}  = \gamma U'(1 + \beta^2) \label{AL2b}
\end{equation}
and

\begin{equation}
{\bm P_{e2}} = 2 \gamma {\bm v} U'/c^2, \label{AL1b}
\end{equation}
while, in the case of a line charge oriented perpendicularly to
the direction of ${\bm v}$,

\begin{equation}
E_{e3}  = \gamma U' \label{AL4}
\end{equation}
and

\begin{equation}
{\bm P_{e3}} = \gamma {\bm v} U'/c^2, \label{AL3}
\end{equation}
which only incidentally, due to the particular choice of the
distribution, behaves as a four-vector.

Henry Poincar\'e solved the problem of the lack of covariance
shown in Eq. (\ref{AL2}) and Eq. (\ref{AL1}) by introducing, in
the electron model,  energies and momenta of non-electromagnetic
nature \cite{POIN}. These are actually due to non-electomagnetic
interactions which keep the electron together. By doing so he
strongly related the covariance of energy and momentum with the
stability of the system:  the electomagnetic energy-momentum pair
alone is not a four-vector, but the total energy-momenum pair,
accounting for the non-electromagnetic interaction, is a regular
four-quantity.

In 1922, Enrico Fermi developed an original, early relativistic
approach to the $4/3$ problem \cite{FERM}; about forty years later
a redefinition of the energy-momentum pair related to Fermi's work
was proposed by Rohrlich \cite{ROHL}, which leaves untouched the
total energy-momentum vector, but splits it into electromagnetic
and non-electromagnetic contribution in such a way that covariance
is granted for both the electromagnetic and the
non-electromagnetic part of the energy-momentum pair.

It is possible to show \cite{TEUK}, \cite{GRIF} that the
treatments by Poincar\'e and Rohrlich are not in contradiction.

Nevertheless, the approach by Rohrlich \cite{ROHL} was sometimes
taken (see e.g. \cite{ERR1}) as the proof that stability and
covariance are unrelated matters since, upon redefinition, the
electromagnetic part alone is a four-vector.

We will show here, that such a conclusion is incorrect. The
stability of the system is related to the covariance of the total
energy-momentum vector, according to the original work by
Poincar\'e: the redefinition  procedure mentioned above is indeed
acceptable only in the case one is interested in the total
energy-momentum vector of a stable system (i.e. a system whose
constituents are and stay at rest in a particularly chosen frame),
and not in the separate electromagnetic and non-electromagnetic
part. Only in that case the arbitrariness included in the
recombination of these two contributions does not affect the
equation of motion for the system (which deals, in fact, with the
total energy-momentum vector).

Some time ago, we were addressing the problem of describing the
transverse self-fields originating within an ultrarelativistic
electron bunch moving in a fixed trajectory \cite{OURS}.

This is a particularly relevant problem in modern particle
accelerator physics, in view of the need for very high-peak
current, low emittance beams to be used, for example, in
self-amplified spontaneous emission (SASE)-free-electron lasers
operating in the x-ray regime (see for example \cite{SAS1},
\cite{SAS2}): in fact, the good quality of the beam may be spoiled
by self-interaction occurring within the bunch.

Besides practical relevance (which stresses how, after one hundred
years, pure academical problems become relevant also to applied
physics), an electron bunch is also a very good example of an
unstable system subject to purely electromagnetic interactions.
For such a system, the total energy and the total momentum in any
frame, are just of electrodynamical nature. We will show that
(according to our previous statement about the relation between
stability and covariance) there is no way, in this case, to define
the total energy-momentum pair in a covariant way. In fact,  in
contrast to what happens for stable systems, there is no way to
describe the evolution of an unstable system without the knowledge
of the (electromagnetic) field theory  governing the
(self-)interactions between its constituents.

\section{\label{prob}A PARADOX AND ITS SOLUTION}

Let us consider a short electron bunch moving, in a given
laboratory frame, in a circular orbit. We can simplify the
description of this system accounting only for two electrons which
will represent the head and the tail of our bunch.

Imagine that the two particles are moving, initially with the same
Lorentz factor $\gamma \gg 1$, in a circular orbit of radius $R$,
and separated by a (curvilinear) distance $\Delta s \ll
R/\gamma^3$.

In this situation the two electrons are near enough to be
considered travelling with the same velocity vector: indeed it can
be shown \cite{SAL1}, that they radiate as a single particle of
charge $2e$ ($e$ being the electron charge) up to frequencies much
above the synchrotron radiation critical frequency (note that,
from a quantitative viewpoint, the expression "much above" is
trivially connected to "how much" $\Delta s \ll R/\gamma^3$). The
requirements specified before consist, from a geometrical
viewpoint, in assuming that, at the beginning of the evolution,
the two particles world-lines are very close: actually,
considering our resolution in space equal to $R/\gamma^3$, they
initially coincide.

This assumption justifies the presence of an inertial frame in
which both particles are, with good approximation, at rest during
the initial part of their evolution. We will refer to it simply as
the rest frame. A quantitative definition of the initial part of
the evolution may be given when a choice is made about close to
zero are the velocities of the particles in the rest frame. Note
that the existence of the rest frame is central for our study
because, referring to it, one can easily analyze the energy and
momentum of the system constituted by the two particles together
with their electromagnetic fields.

By means of a Lorentz transformation, then, we can recover the
same quantities in the laboratory frame.

Starting with the study in the rest frame we will refer,
separately, to mechanical and electromagnetic quantities.

Obviously, in the rest frame, the mechanical momentum of the
system, ${\bm P'_{ne}}$, is zero, and the mechanical energy,
$E'_{ne}$, is just equal to $2m c^2$, where $m$ is the electron
rest mass.

The study of the electromagnetic contributions to energy and
momentum is also trivial.
Since the electrons are at rest they produce electric field only.
Therefore the Poynting vector vanishes and ${\bm P'_{e}}=0$.  On
the other hand, $E_{e}$ is given, simply, by the work $U'$ done
against the field to bring the two particles together
(quasistatically) from a situation in which they are separated by
an infinite distance.

By doing so, of course, we are neglecting, in both ${\bm P'_{e}}$
and $E'_{e}$, the contributions from the acceleration
(self-)fields generated by the system.

This approximation is justified by the fact that we are discussing
the asymptotic behavior for the two particles separated by a very
small distance: then, as it will be clear from Eq.
(\ref{eqmotalt}) and Eq. (\ref{totselfem}), we may assume that the
acceleration field contribution are unimportant, when compared
with the Coulomb one. In fact acceleration effects saturate in the
asymptotic limit of small distance between the two particles
\cite{SAL1}, while Coulomb ones are singular; once again it must
be clear that we are discussing the asymptotic case for small
distance between the two particles. Therefore we have:

\begin{equation}
E'_{e} = U' = e^2/(4\pi\epsilon_0 \gamma \Delta s) \label{prima}
\end{equation}
and

\begin{equation}
{\bm P'_{e}} = 0. \label{seconda}
\end{equation}
Summing up the electromagnetic and mechanical contributions one
gets the total energy and momentum for the system:

\begin{equation}
E'_{tot}=E'_{ne}+E'_{e}=2mc^2+U' \label{enrest}
\end{equation}
and

\begin{equation}
{\bm P'_{tot}} = {\bm P'_{ne}} + {\bm P'_{e}}=0. \label{momrest}
\end{equation}
As already said one may, now, use a Lorentz transformation in
order to calculate these quantities in the laboratory frame.
Again, since we are interested at the beginning of the evolution,
it follows from our assumptions that the two particles evolve with
the same four-velocity vector. Therefore a direction of motion
(which we will designate with z) is well defined for the system in
the laboratory frame and the Lorentz transformation from the rest
frame is, indeed, a simple boost in the $-z$ direction (note that
a good definition of the $z$ direction is equivalent to a good
definition of the rest frame). We will represent this boost with a
matrix with components $\Lambda^{\mu}_{\ \nu}$ with $\mu, \nu =
0... 3$ (where the third component corresponds to the $z$
direction):

\begin{equation}
\Lambda^{\mu}_{\ \nu} = \left( \begin{array}{cccc}
\gamma        &  0  &  0  &  \beta \gamma \\
   0          &  1  &  0  &        0      \\
   0          &  0  &  1  &        0      \\
\beta \gamma  &  0  &  0  &      \gamma   \\
\end{array} \right).
\label{lor}
\end{equation}
The use one makes of $\Lambda^{\mu}_{\ \nu}$ is a critical point
in our derivation. If one (erroneously) assumes that energy and
momentum constitute a four-vector, then he gets,
straightforwardly:

\begin{equation}
(E_{tot}/c, {\bm P_{tot}})^{\mu} = \Lambda^{\mu}_{\ \nu}
(E'_{tot}/c, {\bm P'_{tot}})^{\nu} \label{Lnetot}
\end{equation}
and, therefore,

\begin{equation}
{E_{tot}/c} = \gamma\left(2mc^2 + U'\right),
 \label{totenalt}
\end{equation}
and

\begin{equation}
{P_{tot}} = \gamma\left(2m + U'/c^2 \right)\beta c~,
\label{totmomalt}
\end{equation}
where $P_{tot}$ is a scalar quantity, since we understand that
${\bm P_{tot}}$ is oriented along the $z$ direction.

We can now project the equation of motion, ${\bm F_{syst}} = d
{\bm P_{tot}} / dt$, onto the transverse direction (perpendicular
to $z$ and lying on the orbital plane) thus getting, within our
approximations:

\begin{equation}
F_\mathrm{\bot syst} = 2 e B \beta c +  {e^2  \over {4 \pi
\varepsilon_\mathrm{0} R \Delta s}} \label{eqmotalt}
\end{equation}
As already mentioned in Section \ref{INTRO} (with in mind the
applications in the physics of particle accelerators and the one
of SASE-FELs, see \cite{SAS1}, \cite{SAS2}) we addressed the
description of the transverse self-fields originating within an
electron bunch moving in a circle in \cite{OURS}. In that paper an
approach has been proposed which involves purely electrodynamical
considerations, based on the retarded Green function solution of
Maxwell equations.

In particular, in part of \cite{OURS} we treated the case of two
particles separated by a distance $\Delta s$ (non necessarily much
smaller than $R/\gamma^3$), moving rigidly in a circle (see Fig.
\ref{FIG1}) of radius $R$. Our results disagree with Eq.
(\ref{eqmotalt}).

Let us briefly justify the latter statement. The total transverse
force (orthogonal to its velocity and lying on the orbital plane)
felt by the head electron and due to the tail electron source
turned out to be (see, again \cite{OURS}):

\begin{figure}
\includegraphics*[width=84mm]{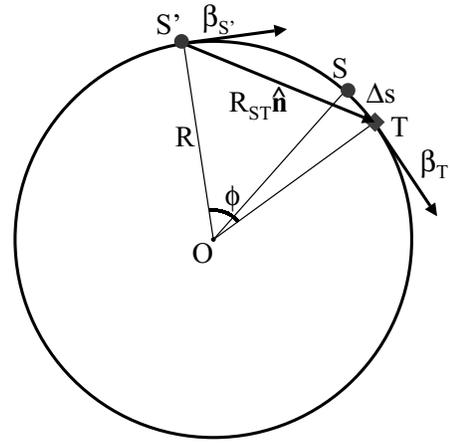}
\caption{\label{FIG1} Geometry for the two-particle system in the
steady state situation, with the test particle ahead of the
source. Here T is the present position of the test particle, S is
the present position of the source, while S' indicates the
retarded position of the source.}
\end{figure}

\begin{equation}
{F_\mathrm{\bot}} \simeq {e^2 \gamma^3 \over {4 \pi
\varepsilon_\mathrm{0} R^2}} \Phi(\hat{\phi})~, \label{Totexp}
\end{equation}
where $\Phi$ is defined by

\begin{equation}
\Phi(\hat{\phi}) = {2+\hat{\phi}^4/8 \over{\hat{\phi}
(1+\hat{\phi}^2/4)^3}} ~,\label{Phi}
\end{equation}
Here $\hat \phi$ is the retarded angle $\phi$ (which expresses the
angular distance between the retarded position of the source the
present position of the test electron, see Fig. \ref{FIG1},
normalized to the synchrotron radiation formation angle at the
critical frequency, $1/\gamma$, i.e. $\hat \phi = \gamma \phi$.
Eq. (\ref{Phi}) is completely independent of the parameters of the
system.

It is straightforward to study the asymptotic behaviors of $\Phi$.
In order to do so, just remember that the retardation condition
linking $\Delta s$ and $\phi$ is given by (see \cite{OURS},
\cite{SAL1}):

\begin{equation}
\Delta s = R\phi -2\beta R\sin{\phi\over{2}}~, \label{retsteady}
\end{equation}
or by its approximated form

\begin{equation}
\Delta s = (1-\beta)R\phi + {R\phi^3\over{24}}~.
\label{retsteadyappr}
\end{equation}
It is now evident that $\Phi(\Delta \hat{s}) \rightarrow
1/(3\Delta \hat{s})$ when $\hat{\phi} \gg 1$ and $\Phi(\Delta
\hat{s}) \rightarrow 1/(\Delta \hat{s})$ when $\hat{\phi} \ll 1$,
having introduced the normalized quantity $\Delta \hat{s} =
(\gamma^3/R) \Delta s$. This normalization choice is, again,
linked with the fact that the critical synchrotron radiation
wavelength, $R/\gamma^3$, is also the minimal characteristic
distance of our system: as we said before, two particles nearer
than such a distance can be considered as a single one radiating,
up to the critical frequency, with charge $2e$ (see \cite{OURS},
\cite{SAL1}).

\begin{figure}
\includegraphics*[width=90mm]{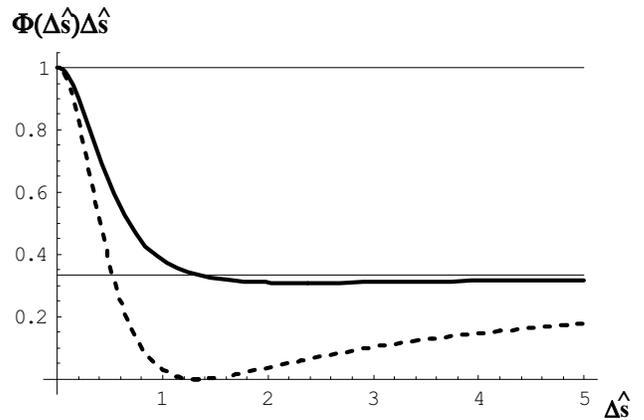}
\caption{\label{FIG2} Plot of $\Phi(\Delta \hat{s})\Delta \hat{s}$
(solid line) and comparison with the asymptotic values, $1$ and
$1/3$. The dashed line shows the contribution from the
acceleration field, $\Phi_\mathrm{R}(\Delta \hat{s})\Delta
\hat{s}$, alone.}
\end{figure}

The asymptotic behavior above suggests to study the function
$\Phi(\Delta \hat{s})\Delta \hat{s}$. We plotted such a function
in Fig. \ref{FIG2} (and the contribution from the acceleration
field alone) for values of $\Delta \hat{s}$ running from 0 to 5.

As it is seen from the figure, the contribution from the velocity
field is not important in the asymptotic limit for particles very
nearby (our case) or very far away. When, in particular, $\Delta s
\ll R\gamma^3$ we can approximate Eq. (\ref{Totexp}) by

\begin{equation}
{F_\mathrm{\bot}} \simeq {e^2  \over {4 \pi \varepsilon_\mathrm{0}
R \Delta s}}~. \label{forzasint}
\end{equation}

\begin{figure}
\includegraphics*[width=84mm]{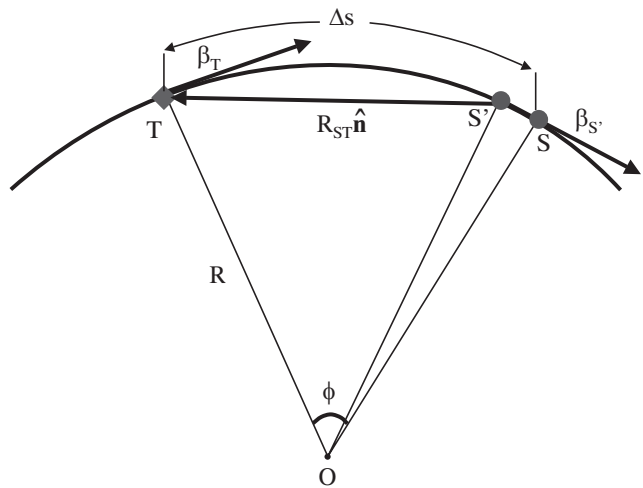}
\caption{\label{FIG3} Geometry for the two-particle system in the
steady state situation, with the source particle ahead of the test
one. Here T is the present position of the test particle, S is the
present position of the source, while S' indicates the retarded
position of the source.}
\end{figure}
On the other hand, as regards the force felt by the tail particle
(see Fig. \ref{FIG3}), it is easily seen that (see \cite{OURS},
\cite{SAL1}) the test electron, which now is the tail particle,
"runs against" the electromagnetic signal emitted by the source
(while in the previous case it just "runs away" from it).
Therefore the relative velocity between the signal and the test
electron is equal to $(1+\beta)c$ (instead of $(1-\beta)c$ in the
other situation). Hence the retardation condition reads

\begin{equation}
\Delta s = R\phi + 2\beta R\sin{\phi\over{2}},
\label{retcondhtexact}
\end{equation}
or, solved for $\phi$ in its approximated form,

\begin{equation}
\phi \simeq {\Delta s\over{R(1+\beta)}}~. \label{retcondht}
\end{equation}
In this situation, $\bm{\beta_\mathrm{S}}$ is almost parallel (and
equal) to $\bm{\beta_\mathrm{T}}$ and antiparallel to $\hat{
\bm{n}}$ (which is the unit vector oriented as the line connecting
the retarded source to the present test particle): it turns out
that the only important contribution to the transverse force
results from the acceleration field and reads:

\begin{equation}
{F_\mathrm{\bot}} \simeq {e^2  \over {4 \pi \varepsilon_\mathrm{0}
R \Delta s}}~.
 \label{Fperpht}
\end{equation}
In the case under study, since $\Delta s \ll R/\gamma^3$, the
total self-force acting on the system is given by the sum of Eq.
(\ref{forzasint}) and Eq. (\ref{Fperpht}):

\begin{equation}
{F_\mathrm{\bot}} \simeq 2{e^2  \over {4 \pi
\varepsilon_\mathrm{0} R \Delta s}}~, \label{totselfem}
\end{equation}
which is in disagreement of a factor $2$ with respect to the
self-force term in Eq. (\ref{eqmotalt}). In other words, had we
erroneously assumed covariance, we would have encountered a
paradox. Again, note that we are treating the asymptotic limit of
a small distance between the two electrons: in this limit we can
neglect the contribution from the acceleration field in the
equation of motion Eq. (\ref{eqmotalt}). In fact this contribution
saturates for small distance between the two particles (see
\cite{SAL1}), while the term missing by comparison between Eq.
(\ref{eqmotalt}) and Eq. (\ref{totselfem}) is singular (and,
therefore, dominating) in the limit when $\Delta s$ goes to zero.

This situation should not be too much surprising for the reader
familiar with the works \cite{ABR1}... \cite{ABR2} which led to
Eq. (\ref{AL2}) and (\ref{AL1}): the derivation of Eq.
(\ref{eqmotalt}) is, in fact, performed under the explicit
assumption that energy and momentum constitute a four-vector.

As we will immediately see, in the case of unstable systems (like
the one we deal with), the use of correct transformation laws for
the electromagnetic stress tensor solves the problem but spoils
the covariance of the energy-momentum pair.

The energy and momentum of an electromagnetic system in the
laboratory frame is given by

\begin{equation}
E = \int T'^{\mu\nu} \Lambda^{0}_{\ \mu} \Lambda^{0}_{\nu} {d
V'\over{\gamma}}
\label{Egen}
\end{equation}
\begin{equation}
P^i = {1\over{c}}\int T'^{\mu\nu} \Lambda^{i}_{\ \mu}
\Lambda^{0}_{\ \nu} {d V'\over{\gamma}},
\label{Pgen}
\end{equation}
where $T^{\mu \nu}$ are the components (in the rest frame) of the
electromagnetic stress tensor of the system, which contains all
the information about the (electromagnetic) field theory governing
the interactions between the particles. The process of lowering
and raising indexes is governed in the usual way by the metric
tensor. Here the latin index $i$ runs from 1 to 3 and, as already
said, the quantities with prime refer to the rest frame. In our
case we will consider the only important component, i.e. the third
(along $z$).

Note that the integrals in Eq. (\ref{Egen}) and Eq. (\ref{Pgen})
include both a single-particle term and an interaction term
(compare also \cite{GRIF}). Here we are interested in the
interaction term alone: in fact we will treat the (trivial)
mechanical contributions to the energy-momentum pair separately
and, once again, we will neglect the acceleration-field
contributions. Therefore, in the following, we will understand
that $T'$ refers to the interaction term alone.

Then, since the mechanical energy-momentum pair of a single
particle is a 4-vector, one gets:

\begin{equation}
(E_{ne}/c,P_{ne})^{\mu} = \Lambda^{\mu}_{\ \nu}
(E'_{ne}/c,P'_{ne})^{\nu}~, \label{Lne}
\end{equation}
hence

\begin{equation}
E_{ne} = 2 \gamma m c^2 \label{euno}
\end{equation}
and

\begin{equation}
P_{ne} = 2 \beta \gamma m c . \label{edue}
\end{equation}
while

\begin{equation}
E_e =  \beta^2 \gamma \int T'^{33} dV'+ \gamma U' , \label{Le}
\end{equation}
\begin{equation}
P_e = \gamma \beta U'/c + \gamma \beta/c \int T'^{33} dV',
\label{LP}
\end{equation}
We should note, here (but this is a valid methodological remark
also as regards the previous, incorrect approach), that the
particles are subject to a long-range interaction (the
electromagnetic interaction) and, therefore, a covariant
definition of the total energy and momentum is not straightforward
even when one is considering the two particles alone, without
including (as we did, instead) the electromagnetic fields in the
system. Therefore, strictly speaking, one may object that Eq.
(\ref{Lne}) does not make any sense at all.

Indeed, if the interaction occurred at a single point in
space-time (short-range scattering case), the particle velocities
would have been constant, in the view of any inertial observer,
before and after the scattering took place. Then if two observers
related by a Lorentz boost compared their judgments about the
particles velocities, they would have found that these are linked
by a Lorentz transformation, the same which transforms from one
observer to the other. Nevertheless, this is a particular case. We
must remember that, in general, according to the Theory of
Relativity, the concept of simultaneity depends on the observer.
Therefore, when (as in the situation under study) one deals with a
long-range interaction, the velocities of the particles, in the
judgment of the two observers above, are not related by a Lorentz
transformation anymore (see \cite{TRUM}): the objection about the
correct definition of the mechanical energy-momentum vector
follows directly from this observation. A good definition of the
energy-momentum vector of the two-particle system alone (without
their fields!) can, in fact, be recovered using more subtle
geometrical methods only, like a regular correlated representation
of the world lines (see \cite{TRUM}).

However this objection does not concern us, here, since, as
already implicit in the definition of the $z$ direction, we
discuss about that region of space-time in which the two particles
world lines are very close and are roughly characterized by the
same Lorentz factor, that is, once again, at the beginning of the
evolution.


Then we can use Eq. (\ref{Lne})... Eq. (\ref{LP}) to get, by
summation, the total energy and momentum of the system in the
laboratory frame and in the direction of motion (see also
\cite{JACK} and \cite{MOLL}):

\begin{equation}
{E_{tot}} = \gamma\left(2mc^2 + U' + \beta^2\int T'^{33}~
dV'\right),
 \label{toten}
\end{equation}
and

\begin{equation}
{P_{tot}} = \gamma\left(2m + U'/c^2 + {1\over{c^2}} \int T'^{33}~
dV'\right)\beta c~. \label{totmom}
\end{equation}
Note that Eq. (\ref{toten}) and Eq. (\ref{totmom}) can be used to
obtain Eq. (\ref{AL2}) and Eq. (\ref{AL1}), as well as Eq.
(\ref{AL2b})... Eq. (\ref{AL3}): different distributions of charge
give different expressions for the electromagnetic stress tensor
and for the electromagnetic energy.

In our case of two electrons we already know the explicit
expression for  $U'$. In fact we remind that, as has already been
said, the electromagnetic interaction energy is simply given by
the work done against the field to bring the two particles
together, quasistatically, from a situation in which they are
separated by an infinite distance:

\begin{equation}
U' \simeq {e^2  \over {4 \pi \varepsilon_\mathrm{0} \gamma \Delta
s}}~.
 \label{U}
\end{equation}
On the other hand it is easy to calculate  $T'^\mathrm{33}$ (in
the rest frame, since we need to integrate over $V'$). To this
purpose we remind that, in the rest frame (and at short distance
$\gamma \Delta s \ll R/ \gamma^2$, so that the acceleration field
contributions are unimportant), the space-space components of the
total (comprehensive of both single particle and interaction part)
symmetric energy-momentum tensor read (see \cite{JACK}):

\begin{equation}
T'^{ij}_{tot}= -\varepsilon_0(E'_{i}E'_{j}- \delta_{ij} E'^2/2)~,
\label{stress}
\end{equation}
where, here, \textit{i}, \textit{j} = 1.. 3. 
The discussion above shows that, for us, the only interesting
component is $T'_{33}$. It can be proven that the interaction part
alone is just

\begin{equation}
\int T'^{33} ~dV'= U' ~.\label{stressequU}
\end{equation}
Note that Eq. (\ref{stressequU}) describes also the case of a
charged line distribution oriented in the direction of motion (in
the case of a charged line, of course, the single-particle term is
not present at all).

The equations for the energy and momentum of the system in the
laboratory frame now read:

\begin{equation}
{E_{tot}} = \gamma\left[2mc^2 + U'(1+\beta^2)\right]
\label{lasttoten}
\end{equation}
and

\begin{equation}
{P_{tot}} = \gamma\left(2m + 2U'/c^2\right)\beta c~,
\label{lasttotmom}
\end{equation}
whose electromagnetic parts are the same of Eq. (\ref{AL2b}) and
Eq. (\ref{AL1b}).

From the transverse component of the equation of motion for the
system one gets

\begin{equation}
F_\mathrm{\bot syst} \simeq 2 e B \beta c + 2 {e^2  \over {4 \pi
\varepsilon_\mathrm{0} R \Delta s}} \label{eqmot}
\end{equation}

Eq. (\ref{eqmot}) is now in perfect agreement with our result in
Eq. (\ref{totselfem}).

Both the terms on the right hand side of Eq. (\ref{eqmot}) are
centripetal as well as in Eq. (\ref{eqmotalt}) (although, of
course, Eq. (\ref{eqmotalt}) and Eq. (\ref{eqmot}) are in
disagreement as concerns the magnitude of the self-interaction
term); the first describes the motion of the system under a
magnetic field, while the second is linked with the presence, in
the system, of electromagnetic fields: an extra centripetal
(external) force is needed, if one wants to keep the system moving
in a circle of radius $R$, compensating for the centrifugal
self-field contributions calculated in \cite{OURS}.

\section{\label{sec:paral} DISCUSSION}

In order to reach the agreement between Eq. (\ref{eqmot}) and Eq.
(\ref{totselfem}), one has to give up the covariance of the
energy-momentum pair of our system, as it is seen directly from
Eq. (\ref{lasttoten}) and Eq. (\ref{lasttotmom}).

We can sum up the discussion in Section \ref{prob} by saying that
the assumption of covariance for the transformation of the
energy-momentum pair of an unstable system leads to a paradox.
Such a paradox can be solved introducing the correct
transformation laws for the energy-momentum tensor. In this way,
it is seen that the energy-momentum pair for an unstable system
(particles and electromagnetic field) is not a four-vector. We
already mentioned in Section \ref{INTRO}, that the non-covariant
character of energy and momentum is also present when one
discusses some classical problems which involve the relativistic
dynamics of a charged particles stable system. Nevertheless, in
these cases, covariance can be always restored by introducing
non-electromagnetic stresses which keep the system together or,
equivalently, by redefinition (see \cite{ROHL}) of the
energy-momentum pair. In particular we can refer to the $4/3$
problem in the classical electron model (as done above in Section
\ref{INTRO}) but also to other problems like, for example,  the
explanation of the Trouton-Noble paradox, which has been treated
extensively in literature (see \cite{TEUK}, \cite{TRA1}..
\cite{BUTL}) all over the last century.

As we already said in Section \ref{INTRO}, there is one major
difference with respect to our case, though: our system, in
contrast with the latter ones, is, by nature, unstable; there is
no reference frame such that its components are and stay at rest.
This fact leads to a major difference in the treatment of the
total energy and momentum.

In the case of a stable system, the total energy and momentum of
the system (including non-electromagnetic binding forces)
constitute a four-vector, a well-defined geometrical entity.

On the contrary, in the case of an unstable system, this pair of
quantities has no geometrical meaning, although it is possible to
give, of course, separate definitions of total energy and momentum
in the judgment of any observer.

In the situation discussed in Section \ref{prob} this is a direct
consequence of the fact that we deal with a fully electrodynamical
system and there is no way to introduce, in a straightforward way,
an analogue of Poincar\'e stresses. As a result we must conclude
that stability of the system and covariance of the energy-momentum
pair are bound together.

Let us discuss the latter statements in detail, starting with a
review of well-known arguments for stable systems.

Stable systems are characterized by a zero (total self-)
four-force density. When the four-force density can be derived
from an energy-momentum tensor $T^{\mu \nu}$, the latter property
is equivalent to:

\begin{equation}
T^{\mu\nu}_{\ \ \ ;\mu} = 0 ~,\label{diverg}
\end{equation}
which is the requirement for a zero-divergency energy-momentum
tensor.

However, Eq. (\ref{diverg}) refers to the total energy-momentum
tensor, while the electromagnetic part of it is not divergenceless
at all (its divergence is, simply, the Lorentz four-force
density).

On the other hand, since stability is characterized by a zero
total four-force density, non-electromagnetic stresses must be
present (Poincar\'e stresses), which balance the Lorentz
four-force density, thus insuring stability for the system.
Poincar\'e stresses also insure covariance for the energy-momentum
pair which is, therefore, a well-defined four-vector: to prove
this, one can remember the definition of the total energy-momentum
pair (see \cite{JACK}, \cite{TEUK}):

\begin{equation}
P^\mu = {1\over{c}} \int_\sigma T^{\mu\nu} d\sigma_\nu
\label{enmom}
\end{equation}
where the integration is carried out over any hypersurface at
$t=constant$ (actually $\sigma$ may be, more generally, any
spacelike surface, see \cite{TEUK}) for any inertial observer.

It can be easily proved (see \cite{TEUK}) that Eq. (\ref{enmom})
is independent from the choice of the integration hypesurface.
Such a proof is based on Eq. (\ref{diverg}).

The choice of different inertial observers is equivalent to the
choice, on the space-time manifold, of different time-like unit
vectors. The different families of hypersurfaces orthogonal to
these vectors represent the physical space at a certain time in
the judgement of different observers. From the independence of the
definition in Eq. (\ref{enmom}) of the choice of the integration
hypersurface follows, therefore, the independence of $P^\mu$ on
the reference frame used to evaluate it, and this constitutes the
proof that $P^\mu$ is a well-defined four-vector.

Since $P^{\mu}$ is independent from the choice of the integration
hypersurface, one is free to choose the one which helps better in
solving problems. Historically, two choices have been used in
explaining, for example, the Trouton-Noble paradox. The first (see
\cite{TEUK}) consists in considering the surface at $t=constant$
for any observer. This leads to the usual expressions for the
electromagnetic energy and momentum in a given frame:

\begin{equation}
{E_{e}\over{c}}=P^{0}_{e} = {1\over{2 c}} \int \left(\epsilon_0
{\bm E}^2+ {{\bm B}^2\over{\mu_0}}\right) dV \label{PO}
\end{equation}
and

\begin{equation}
{\bm P_e} = {1\over{\mu_0 c^2}} \int \left({\bm E} \times {\bm
B}\right) dV, \label{normale}
\end{equation}
where $\mu_0$ is the free space permeability.

While Eq. (\ref{PO}) and Eq. (\ref{normale}) do not constitute a
four-vector, one can straightforwardly solve the problem of the
lack of covariance by introducing Poincar\'e stresses.

The second choice consists in selecting $t=constant$ in the rest
frame of the system:

\begin{equation}
{E_{e}\over{c}}=P^{0}_{e} = {1\over{c}}\int \left[{1\over{2}}
\left(\epsilon_0 {\bm E}^2+{{\bm B}^2\over{\mu_0}}\right) - {\bm
v}\cdot {{\bm E} \times {\bm B} \over{\mu_0 c^2}}\right] dV
\label{speciale1}
\end{equation}
and

\begin{eqnarray}
{\bm P_e} = {\gamma \over{c}} \int \Big[{{\bm E} \times {\bm
B}\over{\mu_0 c}} - {{\bm v}\over{2}} \cdot \left(\epsilon_0 {\bm
E}^2+{{\bm B}^2\over{\mu_0}}\right)  + \nonumber\\&&\cr +
\epsilon_0({\bm v}\cdot {\bm E}){\bm E} + ({\bm v}\cdot {\bm
B}){{\bm B}\over{\mu_0}}\Big] dV &&~ \label{speciale}
\end{eqnarray}
In this case it can be easily shown (see \cite{ROHL}, \cite{TEUK})
that the electromagnetic part of the total energy-momentum pair
is, actually, a four-vector. This is, in fact, the same
redefinition of the four-momentum that Rohrlich used to deal with
the electron problem \cite{ROHL}. One can easily check (see
\cite{TEUK}) that, in this case, the non-electromagnetic part of
the total energy-momentum pair is zero.

All this illustrates the well-known fact that the introduction of
Poincar\'e stresses or the Rohrlich redefinition of energy and
momentum are, in fact, equivalent in essence. The choice of the
integration hypersurface is a matter of taste for stable systems,
since the only important quantity from a geometrical viewpoint is
the total energy-momentum four-vector, given by the sum of the
electromagnetic and the non-electromagnetic part; this sum, by
definition, is independent from such a choice. In other words,
different choices of the hypersurface just split the same, total
quantity onto two parts (electromagnetic and non-electromagnetic)
in two distinct ways but, as quoted from \cite{TEUK} : "The split
into electromagnetic and non-electromagnetic parts is quite
arbitrary".

Our point is that this situation is completely different in the
case of unstable systems, where only electromagnetic forces are
present and Poincar\'e stresses are not.

In the case of unstable systems there is no way one can define the
total energy-momentum four-vector. This statement is justified,
from a mathematical viewpoint, by the fact that

\begin{equation}
T_{tot\ ;\mu}^{\mu\nu} = T_{e\ \ \ ;\mu}^{\mu\nu} \ne 0
~;\label{divergno}
\end{equation}
in fact, from the previous discussion, we know that divergenceless
is an essential ingredient for the independence of the total
energy-momentum pair from the choice of the integration
hypersurface.

Note that Eq. (\ref{divergno}) means that there is a non-vanishing
four-force density field over the space-time. In the case
discussed in Section \ref{prob} we gave a practical example of the
latter statements.

Of course, also for unstable systems, we may consider an observer
and find out the energy and momentum of the system with respect to
that observer, but this quantities will not be covariant, nor we
can recover covariance by integrating the energy-momentum tensor
over a suitable hypersurface (as done with stable systems), since
the electromagnetic energy-momentum pair (which coincides now with
the total energy-momentum pair) would change Eq. (\ref{lasttoten})
and Eq. (\ref{lasttotmom}), thus  giving an unphysical result on
the second term of Eq. (\ref{eqmot}), when compared with Eq.
(\ref{totselfem}). The latter discrepancy, to the authors view, is
similar to the one encountered in \cite{GRIF}. In that paper a
system composed by two electron is studied too and a comparison is
proposed between energy-derived mass $m_u$ (electrostatic energy
divided by $c^2$), momentum derived mass $m_p$ (momentum divided
by $\gamma v$, being $v$ the system velocity) and
self-force-derived mass $m_s$ (self-force divided by $\gamma^3 a$,
$a$ being the system acceleration). The authors of \cite{GRIF}
point out that $m_s = m_p \neq m_u$ and that the inequality
between $m_p$ and $m_u$ can be solved by redefining, following
Rohrlich (see \cite{ROHL}) the energy-momentum pair. Nevertheless,
in this case, one is left with a discrepancy between $m_s$ and
$m_u$. This fact is perceived by the authors of \cite{GRIF} as an
unsolved paradox. Actually the derivation of Eq. (\ref{eqmotalt})
(or equivalently, of $m_u$, treated with Rohrlich's method) is
performed under the explicit assumption that energy and momentum
constitute a four-vector which is true only in the case of a
stable system. As a result, by comparing Eq. (\ref{eqmotalt}) and
Eq. (\ref{totselfem}) or, which is the same, $m_u$ (treated by
Rohrlich's method or, equivalently, by introducing Poincar\'e
stresses) with $m_s$, one is comparing quantities which refer to a
stable system with quantities which refer to an unstable one, thus
giving a paradoxical result.

Consider, as a last example, an unstable system formed by
different subsystems initially at rest in a certain frame (as in
the case discussed in Section \ref{prob}). In general, while
dealing with unstable systems, the knowledge of dynamical
quantities for the subsystems cannot bring any information about
the behavior of the system as a whole, unless we have knowledge of
the (electromagnetic) field theory governing the interactions
(which make the system unstable).

In our example of Section \ref{prob}, even if we can measure, in a
certain frame,  the particle velocities when they are far away
from each other (thus no more interacting) and if we know their
rest masses, we cannot say anything about energy and momentum of
the total system without the knowledge of the stress tensor, and
the reason for that is the presence of a non-zero four-force
density field (which we can account for only knowing the stress
tensor, i.e. the interaction theory) on that part of the
four-dimensional manifold on which we want to have information.

Suppose our system was stable (think about an ideal "rope"
responsible for Poincar\'e stresses, or think, instead of the case
of two electrons, about a nuclear fission process in which
electromagnetic fields are, before the fission event, balanced by
the strong interaction). Put ourselves in the laboratory frame,
and imagine that, at a certain moment, for a certain reason (a
collision with a neutron, for example) the Poincar\'e stresses are
not present anymore. Thus the system becomes unstable, it breaks
in two subsystems and the electromagnetic interaction (if you're
thinking about the nucleus imagine we are talking about a charged
ion) takes over. If we wait for enough time the two particles will
get far away from each other and we can consider them no more
interacting.

At this point, the kinetic energy of the particles is equal to the
energy previously stored in the electromagnetic field when the
system was stable: thus we are able to get information about the
total energy-momentum vector of the stable system even if we do
not know anything about the stress tensor and the theory of the
interaction between the subsystems. In fact, the sum of the
momenta of the two free particles and the sum of their energies
will give us, respectively, the energy and the momentum of the
stable system, which form a four-vector again.

The reason is, simply, that there is no four-force density field
in that part of the space time on which we want to have
information: from a general viewpoint we can conclude that the
presence of a four-force density, which characterizes unstable
systems, spoils without remedy the covariance of the
energy-momentum pair. The only way to recover such covariance
would be to introduce a balancing four-force density, i.e. to make
the system stable.

In other words, in agreement with Poincar\'e, from the stability
of the system follows the covariance of the system and, vice
versa, from covariance follows stability: the energy-momentum pair
of an unstable system does not constitute a four-vector. This
conclusion may seem, at first glance, of academical importance
only. It has, indeed, very practical consequences in modern
electron beam physics. Energy and momentum of an electron bunch
are physically measurable quantities and an electron bunch itself
is a practical example of an unstable system.  Nowadays,
technology allows the production of ultra high-brightness, intense
electron beams (to be used, for example, in self-amplified
spontaneous emission (SASE)-free-electron lasers operating in the
x-ray regime). The production of such bunches is one of the most
challenging activities for particle accelerators physicists. The
description of these systems would be completely incorrect without
accounting for self-interactions in the right way. We can conclude
that, for example, simulation codes (as well as analytical
considerations: actually, as already reminded, we wrote this paper
after studying the self-interactions within an electron bunch)
which rely on the covariance of the energy-momentum pair would
give, $a \ priori$, wrong results which may be immediately
confuted by experimental control. Technological developments often
transform, as in this case, purely methodological issues into very
practical ones.

\section{\label{sec:acknowl}ACKNOWLEDGEMENTS}
The authors acknowledge Reinhard Brinkmann, Georg Hoffstaetter,
Martin Dohlus, Helmut Mais and Evgeni Schneidmiller (DESY),
Yaroslav Derbenev, Rui Li and Yuhong Zhang (Jefferson Lab),
Mikhail Yurkov (JINR), Jan Botman, Jom Luiten and Marnix van der
Wiel (Technische Universiteit Eindhoven) with useful discussions.
Moreover G.G. greatly appreciates financial support from the
Centrum voor Plasmafysica en Stralingstechnologie (CPS, the Dutch
Centre for Plasma Physics and Radiation Technology) and both
financial support and hospitality granted by the Center for
Advanced Studies of Accelerators (CASA) at Thomas Jefferson
National Laboratory during his work on this paper.

\end{document}